\documentclass[prl, reprint]{revtex4-1}
\usepackage{amsmath,amsbsy, amssymb}
\usepackage[dvips]{graphicx}
\usepackage{xcolor}
\usepackage{units}

\DeclareGraphicsExtensions{.eps,.ps}

\usepackage{bm}

\def\Tr#1{{\textrm{ Tr}} \left( #1 \right)}

\def\det#1{\textrm{ Det}( #1 )} 

\def\ber{\begin{eqnarray}}
\def\eer{\end{eqnarray}}
\def\be{\begin{equation}}
\def\ee{\end{equation}}
\def\beno{\begin{equation*}}
\def\eeno{\end{equation*}}
\def\bea{\begin{eqnarray}}
\def\eea{\end{eqnarray}}

\begin{document}

\title{The Plateau-Rayleigh instability in solids is a simple phase separation}

\author{Chen Xuan}
\affiliation{Cavendish Laboratory, University of Cambridge, 19 JJ Thomson Avenue, Cambridge CB3
0HE, United Kingdom}
\author{John Biggins}
\affiliation{Cavendish Laboratory, University of Cambridge, 19 JJ Thomson Avenue, Cambridge CB3
0HE, United Kingdom}
\date{\today}
\begin{abstract}
A long elastic cylinder, radius $a$ and shear-modulus $\mu$,  becomes unstable given sufficient surface tension $\gamma$. We show this instability can be simply understood by considering the energy, $E(\lambda)$, of such a cylinder subject to a homogenous longitudinal stretch $\lambda$. Although $E(\lambda)$ has a unique minimum, if surface tension is sufficient ($\Gamma\equiv\gamma/(a\mu)>\sqrt{32}$) it looses convexity in a finite region. We use a Maxwell construction to show that, if stretched into this region, the cylinder will phase separate into two segments with different stretches $\lambda_1$ and $\lambda_2$. Our model thus explains why the instability has infinite wavelength, and allows us to calculate the instability's sub-critical hysteresis loop (as a function of imposed stretch), showing that instability proceeds with constant amplitude and at constant (positive) tension as the cylinder is stretched between $\lambda_1$ and $\lambda_2$. We use full nonlinear finite-element calculations to verify these predictions, and to characterize the interface between the two phases. Near $\Gamma=\sqrt{32}$ the length of such an interface diverges introducing a new length-scale and allowing us to construct a 1-D effective theory. This treatment yields an analytic expression for the interface itself, revealing its characteristic length grows as $l_{wall}\sim a/\sqrt{\Gamma-\sqrt{32}}$.
\end{abstract}
 \maketitle

Molecules at a condensed phase's surface have fewer neighbors than those in the bulk, so all materials suffer an energy proportional to their exposed area. The resulting tendency to reduce area is familiar in fluids, explaining why droplets are round, taps drip and pond-skaters don't drown. An isolated fluid body always form a sphere under surface tension, but the corresponding problem for a solid body is altogether more subtle, pitting surface tension against bulk elasticity. Here we give a full account of this competition in long solid cylinders. 

In general a surface energy, $\gamma$, can only compete with a bulk elastic shear-modulus, $\mu$, at scales below the elastocapiliary length, $l_{cap}=\gamma/\mu$, which is sub-Angtrom in crystalline materials but reaches up to microns or even millimeters in soft solids such as gels and biological tissues. Elastocapillary distortions \cite{roman2010elasto} thus occur in the realm of the small and soft, including microfluidics  \cite{van2009formation}, the  processing of polymer strands \cite{zuo2005experimental, naraghi2007mechanical} and many aspects of biology  \cite{hazel2005surface, hannezo2011instabilities, dervaux2011shape, ElastocapillaryInstabilityinMitochondrialFission}. In addition to the deformations of soft cylinders considered here \cite{matsuo1992patterns, barriere1996peristaltic, mora2010capillarity,ciarletta2012peristaltic,ciarlettanonlinearjmps, ciarlettanonlinearpre,xuan2016finite}, recent work has highlighted capillarity-driven bending of wet elastic rods and sheets \cite{CharlotteCapillaryOrigami, bico2004adhesion, kim2006capillary, mora2013solid}, elastic modifications to wetting \cite{style2013universal, style2013surface,style2013patterning}, and the inhibitory role of surface tension in elastic creasing \cite{amar2010swelling, haywood2010nucleation,mora2011surface} and cavitation \cite{gent1990cavitation}.




\begin{figure}[h]
\includegraphics[width=0.9\columnwidth]{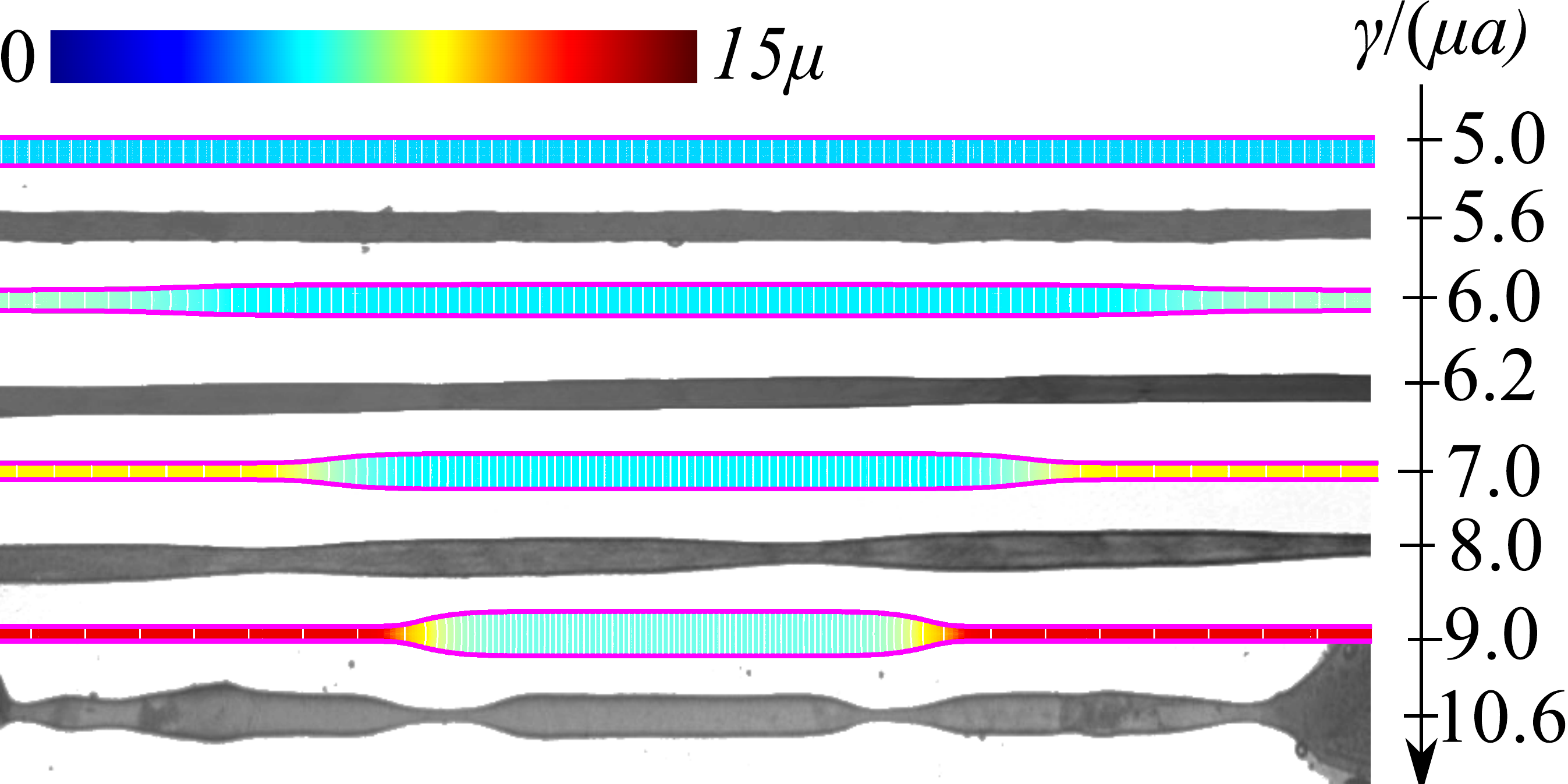}
\caption{Equilibrium shapes of agar-gel \cite{mora2010capillarity} and finite-element cylinders (radius $a$, shear-modulus $\mu$) ordered by their elastocapiliary number $\Gamma=\gamma/(\mu a)$ and showing Plateau-Rayleigh instability for $\Gamma\gtrsim 6$. The finite element instability adopts the longest possible wavelength, and is colored by pressure.}\label{fig:experimental_picture}
\end{figure}

\begin{figure*}
\begin{center}
\includegraphics[width=180mm]{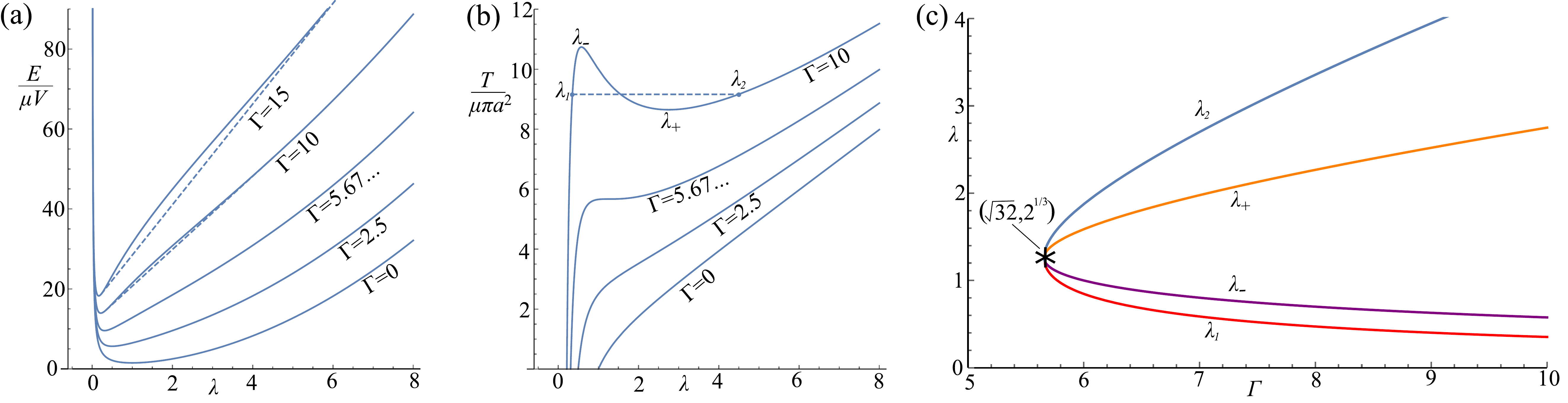}
\caption{Energy (a) and tension (its derivative, b) of an elastic cylinder stretched by $\lambda$ at various elastocapiliary numbers $\Gamma$. As $\Gamma$ increases the minimum-energy/zero-tension stretch moves below one and, for $\Gamma>\sqrt{32}=5.67..$, a region of negative curvature appears in the energy between $\lambda_{-}$ and $\lambda_{+}$, corresponding to an unstable region where tension falls with extension. In (a) the dashed line indicates the energy's common-tangent chord, which spans the concave region meeting the energy at  $\lambda_1<\lambda_{-}$ and $\lambda_2>\lambda_{+}$. The cylinder can pay the lower common-tangent energy by phase-separating into $\lambda_1$ and $\lambda_2$ regions and (dashed line in b) thus stretches from $\lambda_1$ to $\lambda_2$ at constant tension. Fig (c) plots the stretch ratios $\lambda_1$, $\lambda_2$ and $\lambda_{\pm}$ as a function of $\Gamma$.}
\label{energy_plot}
\end{center}
\end{figure*}

In fluid cylinders, surface tension famously causes disintegration into spherical droplets \cite{strutt1879instability}. The driver for this \emph{Plateau-Rayleigh instability} is that an undulation along a cylinder reduces its surface area if its wavelength is longer than its circumference. How the instability proceeds depends in on the fluid \cite{bhat2010formation, pinching},  but this simple geometric origin means, inescapably, that all fluid cylinders ultimately disintegrate. A solid cylinder (radius $a$) is also subject to the same geometric reality, but in this case, the surface energy saved by undulation only outweighs the elastic cost of deforming the cylinder if $l_{cap}\gtrsim a$ \cite{barriere1996peristaltic}, so only thin soft cylinders are unstable. Solid cylinders cannot break into spheres, so the instability produces stable undulating cylinders, fig.\ \ref{fig:experimental_picture}, with a \emph{beads-on-a-string} morphology emerging at high surface tension. Previous groups have observed this instability \cite{matsuo1992patterns, mora2010capillarity} and studied its onset via linear stability theory \cite{barriere1996peristaltic,mora2010capillarity}, showing the first unstable mode has infinite wavelength. More recent work has focussed on stretched cylinders  \cite{ciarletta2012peristaltic,ciarlettanonlinearjmps, ciarlettanonlinearpre}, showing, with linear stability theory, that the instability occurs in a particular stretch-interval, $\lambda_1<\lambda<\lambda_2$, and, via finite element numerics, that, between  $\lambda_1$ and $\lambda_2$, the bead amplitude is essentially constant while bead length falls \cite{ciarlettanonlinearpre}. Here we demonstrate that this long-wavelength constant-amplitude beading arises because the instability is actually a longitudinal phase separation between a less and a more stretched phase.  We are able to simply predict the instability's full high-amplitude behavior, including its hysteresis loop, verify our predictions with finite elements, and identify and analyze the domain walls separating the two phases. 

We consider a long elastic cylinder, radius $a$ and length $L$, that undergoes a displacement $\mathbf{u}$ and consequent deformation gradient $F=I+\nabla \mathbf u$. We model the cylinder as incompressible and neo-Hookean with shear-modulus $\mu$ and surface energy $\gamma$, so the cylinder's total energy is\begin{equation}
E=\gamma A+\int \frac{1}{2}\mu\left[\Tr{F\cdot F^T}\right] d V,
\end{equation}
where $A$ is the cylinder's external area, and incompressibility requires $\det{F}=1$. Previous studies have shown this energy first becomes linearly-unstable to an infinite wavelength perturbation when $\gamma=6 \mu a$  \cite{barriere1996peristaltic,mora2010capillarity}. We first verify that long-wavelength behavior persists far beyond threshold by conducting full finite element calculations at  $\gamma=8 \mu a$ in long aspect ratio cylinders ($L=100 a$, see numerical SI) and observe that, although instability arises with a finite wavelength, it coarsens (movie S1), saving energy, until, as seen in fig.\ \ref{fig:experimental_picture}, the longest possible mode is reached. To understand this, we imagine stretching the cylinder homogeneously to a length $\lambda L$, causing a contraction of its radius to $a/\sqrt{\lambda}$ so that the volume, $V=\pi a^2 L$, is preserved. The cylinder's surface area thus increases to $2 \pi a L \sqrt{\lambda}$, and its deformation gradient is $F=\mathrm{diag}(1/\sqrt{\lambda},1/\sqrt{\lambda},\lambda)$, so its  total energy is
\begin{align}
E(\lambda)&=\pi \mu L a^2 \left(2 \Gamma \sqrt{\lambda}+\frac{1}{2}\left(\lambda^2+\frac{2}{\lambda}\right)\right),
\end{align}
where we have introduced  $\Gamma=\gamma/(\mu a)=l_{cap}/a$, the elastocapillary number for the cylinder. We plot $E(\lambda)$ for several values of $\Gamma$ in fig.\ \ref{energy_plot}a. The elastic part of the energy is a convex function minimized at $\lambda=1$. The surface energy never dominates at  small or large $\lambda$, and the the total energy always has a single minimum, though it moves to $\lambda<1$ as $\Gamma$ increases \cite{mora2013solid}. However, the surface tension term, proportional to $\sqrt{\lambda}$, is a concave function and, when $\Gamma$ is sufficient, it introduces a concave region into the total energy. This concave region ($\frac{d^2 E}{d \lambda^2}<0$) is mechanically unstable since therein the tension,

\begin{figure*}
\begin{center}
\includegraphics[width=180mm]{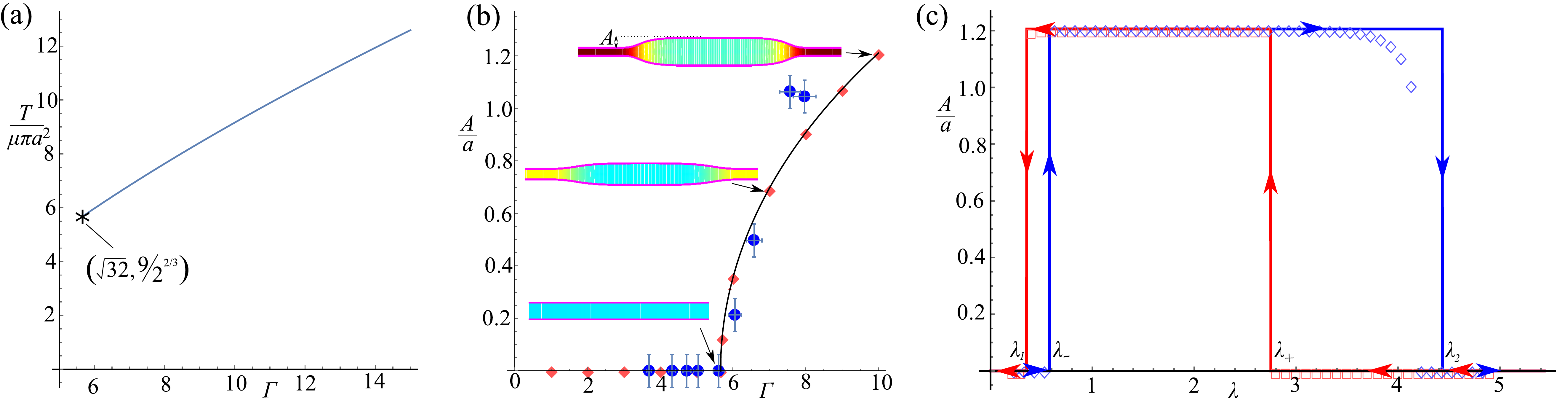}
\caption{Tension (a) and amplitude (b) of surface-tension driven phase-separation in an elastic cylinder as a function of elastocapiliary number $\Gamma$. In (b) we compare theory (solid line), finite elements (red diamonds) and experimental data (blue circles) from \cite{mora2010capillarity}.  (c): Comparison of theoretical (solid lines) and finite element (points) hysteresis loops showing amplitude as a function of imposed stretch at $\Gamma=10$; the constant amplitude is the hallmark of phase separation.}
\label{fig:amp}
\end{center}
\end{figure*}

\begin{equation}
T=\frac{1}{L}\frac{d E}{d \lambda}=\pi \mu  a^2  \left(\frac{\Gamma }{\sqrt{\lambda}}+\lambda-\frac{1}{\lambda^2}\right),
\end{equation}
falls with increasing stretch, as seen in fig.\  \ref{energy_plot}b. The limits of this mechanically unstable region are
\begin{equation}
\lambda_{\pm}=\left[\frac{1}{4}\left(\Gamma\pm\sqrt{\Gamma^2-32}\right)\right]^{2/3},\label{eq:spinodal}
\end{equation}
where $\frac{d^2 E}{d \lambda^2}=0$. The concave region arises when $\Gamma>\sqrt{32}$, first appearing at $\lambda\pm=2^{1/3}$ and spreading out reaching $\lambda_{-}=1$ at $\Gamma=6$, both in agreement with previous linear stability thresholds \cite{mora2010capillarity, ciarlettanonlinearjmps}. In general, given a pair of stretch ratios, $\lambda_1$ and  $\lambda_2$, the cylinder can achieve on average any intermediate stretch $\lambda_1<\lambda<\lambda_2$ by phase separating into a $\lambda_1$ region and a $\lambda_2$ region with appropiate length fractions, at an energy cost that corresponds to the chord connecting $E(\lambda_1)$ and $E(\lambda_2)$.  In the concave region such chords lie below the original energy, so phase-separation saves energy. As understood by Maxwell \cite{maxwell_construction}, the optimal phase-separation arises when the chord is tangent to the energy at both $\lambda_1$ and $\lambda_2$:
\begin{align}
E'(\lambda_1)&=E'(\lambda_2)\label{eq_common_tangent_1}\\
E(\lambda_1)+E'(\lambda_1)(\lambda_2-\lambda_1)&=E(\lambda_2)\label{eq_common_tangent_2}.
\end{align}
The solutions for $\lambda_1$ and $\lambda_2$ as functions of $\Gamma$ are shown in fig.\ \ref{energy_plot}c. As the cylinder is stretched from $\lambda_1$ to $\lambda_2$, it lowers its energy by phase separating onto the common tangent line, shown as a dashed line in fig.\ \ref{energy_plot}a, stretching from $\lambda_1$ to $\lambda_2$ by adjusting the length-fraction of the two phases. Since the resulting effective energy is a straight line, the \emph{tension is constant during the transition} as shown by the dotted line in fig.\ \ref{energy_plot}b. Furthermore, an obvious conclusion of this phase separation approach is that \emph{no amount of surface tension will destabilize a solid cylinder unless it is also stretched to the transition tension} $T=(1/L)E'(\lambda_1)$, plotted in fig.\ \ref{fig:amp}a. Following the common-tangent yields a convex effective energy, so there is no additional instability. 

When it is phase-separated, the difference in final radius between the $\lambda_1$ segment (the ``bead'') and the $\lambda_2$ segment (the ``string'') is simply
\be
A=\frac{a}{\sqrt{\lambda_1}}-\frac{a}{\sqrt{\lambda_2}}.
\ee
This {\emph{amplitude does not change as the cylinder is stretched}} from $\lambda_1$ to $\lambda_2$.  Such constant amplitude beading between two stretch ratios, (movie S2),  has been observed without explanation  in rat sciatic nerve \cite{markin1999biomechanics} and elasto-capiliary finite elements \cite{ciarlettanonlinearpre}.  We compare experimental amplitudes with our finite-element and phase-separation amplitudes in fig.\ \ref{fig:amp}b. The three agree well, with finite elements exactly reproducing the phase-separation amplitudes, verifying our approach. 
\begin{figure*}
\begin{center}
\includegraphics[width=180mm]{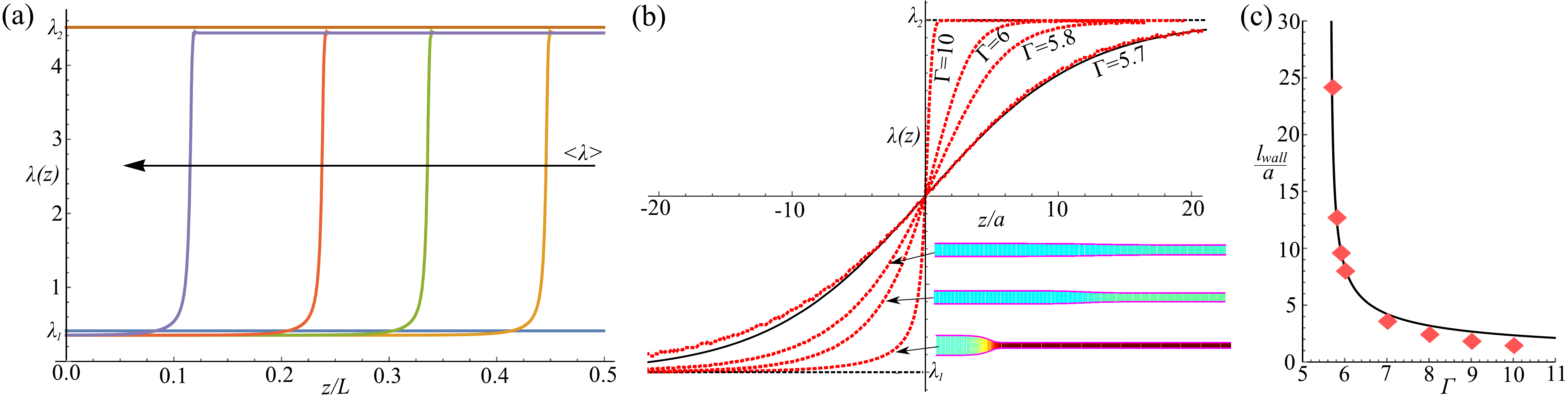}
\caption{Domain walls between separated phases. (a) Finite element calculation of stretch ratio $\lambda(z)$ along a $\Gamma=10$ cylinder during phase separation, showing a $\lambda_1$ and $\lambda_2$ regions separated by a domain wall that moves along the cylinder as the average stretch, $\left<\lambda\right>$, is increased. (b) Finite element domain walls (red dots) in cylinders at various elastocapiliary numbers $\Gamma$.  The domain walls spreads out and becomes more symmetric as $\Gamma\to\sqrt{32}$, agreeing with the black theoretical curve (eqn.\ (\ref{eq:tanh})) by $\Gamma=5.7$. (c) Comparison of theory (curve) and finite-elements (points) for the domain wall length, $l_{wall}$, as a function of $\Gamma$.}
\label{fig:domain_wall}
\end{center}
\end{figure*}

If instability were produced by increasing $\Gamma$ it would be continuous (supercritical) but if it is controlled at fixed $\Gamma$ by stretching from $\lambda_1$ to $\lambda_2$ it will be observed as a discontinuous (subcritical)  with constant amplitude $A(\Gamma)$ throughout. In a stretching cycle the cylinder will be stretched from the energy minimizing contraction $\lambda_{min}$ to $\lambda_1$ homogeneously. Upon passing $\lambda_1$ it would save energy by phase separating, but the cylinder nevertheless remains mechanically stable until it reaches the end of convexity, $\lambda_{-}$, only then  phase separating into $\lambda_1$ and $\lambda_2$. As stretching continues further, the cylinder shifts material from the $\lambda_1$ phase to the $\lambda_2$ phase, causing bead width to fall at constant amplitude, until all the material is at $\lambda_2$, concluding the instability. Similarly in unloading, the instability should occur at $\lambda_{+}$, but then persist down to $\lambda_1$. We thus predict the sub-critical hysteresis loop for a stretch cycle shown in fig.\ \ref{fig:amp}c. Hysteresis  also slightly complicates the tension-extension behavior: we expect the tension to remain on the homogeneous curve in loading until $\lambda_{-}$ then drop to the phase separation value, and remain there until $\lambda_2$. 

Our hysteresis predictions are compared with a finite element stretch-cycle in fig.\ \ref{fig:amp}c, showing good agreement with one discrepancy: the finite-element loading cycle amplitude drops off slightly early. The reason for this is made evident by plotting  (fig.\ \ref{fig:domain_wall}a) the longitudinal stretch ratio $\lambda(z)$ (on the axis of the cylinder) as a function of length along the cylinder, running from the middle of the bead ($z=0$) to the end of the cylinder ($z=L/2$), for various values of the total imposed stretch $<\lambda>$ that traverse the instability. In these plots we see clearly that the cylinder has separated into $\lambda_1$ and $\lambda_2$ regions, but also that they are connected by a transition region---a \emph{domain wall}---with finite length. The energy cost and length-fraction of a domain wall is negligible in an infinite cylinder, but matters in a finite one: upon stretching a finite cylinder  towards $\lambda_2$ the amplitude drops early because the vanishing length of the $\lambda_1$ segment becomes comparable to the domain wall, so the domain-wall raises the phase-separation energy above the homogeneous energy. As expected, in longer cylinders, the instability persists closer to $\lambda_2$. In contrast, at the onset of instability ($\lambda_{-}$) the energy minimizing configuration already requires an extensive length fraction of both phases, so instability proceeds even in modest length cylinders. 

\emph{Around the critical point}, $\Gamma=\sqrt{32}$ and $\lambda=2^{1/3}$, we can solve the phase separation model analytically, and \emph{model the domain wall}. Expanding  (\ref{eq_common_tangent_1}-\ref{eq_common_tangent_2}) in $\epsilon=\Gamma-\sqrt{32}$, we see the leading order behavior of $\lambda_1$ and $\lambda_2$ is
\begin{align}
\lambda_1=2^{1/3}-\frac{1}{\sqrt{3}} 2^{7/12} \sqrt{\epsilon }\mathrm{,\ \  } \lambda_2=2^{1/3}+\frac{1}{\sqrt{3}} 2^{7/12}  \sqrt{\epsilon },
\end{align}
and hence the amplitude and tension are 
\begin{align}
A=a \frac{1}{\sqrt{3}} \sqrt[12]{2} \sqrt{\epsilon } \mathrm{,\ \ \ \ \ \ \ \ } T=\frac{\pi  \mu  a^2} {2^{2/3}}\left(\sqrt{2} \epsilon +9\right).
\end{align}
To model the domain wall we require an energy that penalizes stretch gradients along the cylinder. Such a term could arise from additional elastic shear or additional surface area. We capture both effects by allowing the longitudinal stretch, $\lambda(z)$, to vary slowly along the cylinder's length so a point initially at $(r,z)$ is moved to $(r/\sqrt{\lambda(z)}, \int \lambda(z)d z)$. This introduces a new shear term into the deformation gradient, $F_{rz}=-r \lambda'/(2 \lambda^{3/2})$,  increasing the elastic energy per unit-length to
\begin{equation}
E_{el}=\frac{1}{2}\mu\left(\frac{2}{\lambda^2}+\lambda^2+\frac{a^2 \lambda'^2}{16 \lambda^3}\right)\pi a^2
\end{equation}
and the surface energy per-unit length to
\begin{equation}
E_{s}=\gamma \frac{2 \pi a}{\sqrt{\lambda}}\sqrt{\lambda^2+\frac{a^2 \lambda'^2}{4 \lambda^3}}.
\end{equation}
Finally, we use $T$ as a Lagrange multiplier constraining the average stretch to $\left<\lambda\right>$, leading us to the total energy
\begin{equation}
E=\int E_{s}+E_{el}-T(\lambda-\left<\lambda\right>) d z.
\end{equation}
Near the critical point we can take the above form for $T$ and write  $\lambda=2^{1/3}+\frac{1}{\sqrt{3}} 2^{7/12} \sqrt{\epsilon}h(z)$, where $h(z)$ varies between $\pm1$ through a domain wall. Expanding in $\epsilon$ and keeping only the lowest order terms  $h'$ and $h$ terms yields 
\begin{equation}
E=\!\!\!\int\!\! \mathrm{const}+\frac{17 \pi  a^4 \mu  \epsilon  h'^2}{24\ 2^{5/6}}+\frac{\pi  a^2 \mu  \epsilon ^2}{12 \sqrt[3]{2}} h^2 \left(h^2-2\right)d z,
\end{equation}
where the $\epsilon^2$ term is minimized by $h=\pm1$, while the $\epsilon$ term penalizes gradients. Minimizing $E$ w.r.t\ variations in $h$ requires $17 a^2 h''-4 \sqrt{2} \epsilon  h \left(h^2-1\right)=0$, and hence 
\begin{equation}
h(z)=\tanh \left(\frac{2^{3/4} \sqrt{\epsilon }}{\sqrt{17} a} z\right),\label{eq:tanh}
\end{equation}
an \emph{explicit form for the domain-wall}.  The walls are exponentially localized with a characteristic length  $l_{wall}= 2\times\sqrt{17} a/(2^{3/4}\sqrt{\epsilon})$ that diverges near the critical point. In fig.\ \ref{fig:domain_wall}b we plot $\lambda(z)$ through finite-element domain walls  at various $\Gamma$, showing very good agreement with eqn.\ (\ref{eq:tanh}) near $\Gamma=\sqrt{32}$. In fig.\ \ref{fig:domain_wall}c we show the length of the finite element domain walls (defined as the $z$ interval required for $\tanh(1)\approx0.76$ of the $\lambda$ variation) diverges near the critical point, in accord with our prediction. Although this analysis captures the essential physics, it assumes $z$ displacements are independent of $r$; relaxing this assumption leads to a small decrease in $l_{wall}$ by a factor of $\sqrt{34/33}$ (see theoretical supplement).

In summary, the solid Plateau-Rayleigh instability is a phase separation between regions of different but homogeneous longitudinal stretches connected by exponentially localized domain walls. This structure, reminiscent of elastic necking \cite{antman1973nonuniqueness, ericksen1975equilibrium}, balloon instabilities \cite{gentballoon, doiballoon} and volume transitions in swelling gels \cite{PhysRevLett.40.820, matsuo1992patterns, PhysRevE.68.021801}, provides a picture of the instability that is both simple and complete.  Since phase-separation originates solely in the non-convexity of surface energy, this picture will certainly generalize to more sophisticated elastic constitutive laws, Voigt viscoelasticity and more elaborate extruded prismatic shapes, and, since it naturally produces a high amplitude beads-on-a-string morphology, may underpin the more famous bead/string pattern formed during the breakup of viscoelastic threads \cite{bhat2010formation}. Mastering these instabilities is essential for the better sculpting, spinning and processing of polymer fibers \cite{zuo2005experimental, naraghi2007mechanical}.



\begin{acknowledgments}
{\it Acknowledgements}. We thank S.\ Mora for sharing data and images, and T.\ Tallinen, on whose code our finite element calculations are based. C.X. thanks the China Scholarship Council and the EPSRC for funding.
\end{acknowledgments}


%

\end{document}